\documentclass[apl,aps,preprint]{revtex4-1}

\usepackage{graphicx}

\begin{document}

\title{Strong mobility degradation in ideal graphene nanoribbons due to phonon scattering}

\author{A. Betti}

\email[E-mail: ]{alessandro.betti@iet.unipi.it}

\affiliation{Dipartimento di Ingegneria dell'Informazione: Elettronica, Informatica, Telecomunicazioni,
Universit\`a di Pisa,
Via Caruso 16, I-56122, Pisa, Italy}

\author{G. Fiori}
\affiliation{Dipartimento di Ingegneria dell'Informazione: Elettronica, Informatica, Telecomunicazioni,
Universit\`a di Pisa,
Via Caruso 16, I-56122, Pisa, Italy}

\author{G. Iannaccone}
\affiliation{Dipartimento di Ingegneria dell'Informazione: Elettronica, Informatica, Telecomunicazioni,
Universit\`a di Pisa,
Via Caruso 16, I-56122, Pisa, Italy}

\begin{abstract}


We investigate the low-field phonon-limited 
mobility in armchair graphene nanoribbons (GNRs)
using full-band electron and phonon dispersion relations. 
We show that lateral confinement suppresses the intrinsic mobility of GNRs 
to values typical of common bulk semiconductors, and very far
from the impressive experiments on 2D graphene.
Suspended GNRs with a width of 1~nm exhibit a 
mobility close to 500~cm$^2$/Vs at room temperature, 
whereas if the same GNRs are deposited on HfO$_2$ mobility is further
reduced to about 60~cm$^2$/Vs due to surface phonons. 
We also show the occurrence of polaron formation, leading to band gap 
renormalization of $\approx$118~meV for 1~nm-wide armchair GNRs.

\end{abstract}

\maketitle

\newpage

Understanding the role of phonon 
scattering~\cite{Fratini,Perebeinos,Konar,Fang} is of 
primary importance, since it provides information regarding the ultimate 
{\em intrinsic} mobility limit ($\mu_{in}$) of a material, i.e., when all {\em extrinsic}
scattering sources have been removed. This is especially
important for new materials or nanostructured ones, such as graphene nanoribbons,
where experiments are not fully comparable, given the presence of 
defects and other non idealities.
Recent experiments have found $\mu_{in}$ in suspended graphene close to 
10$^5$~cm$^2$/Vs near room temperature~\cite{Bolotin}, 
but with a sizable degradation after deposition 
on high-{\em k} gate insulators~\cite{Chen}, probably due to 
coupling to the polar modes of the substrate ~\cite{Fratini,Perebeinos,Konar}. 

Very few indications are 
available on mobility degradation in Graphene NanoRibbons (GNRs) 
deposited on different gate insulators~\cite{Liao}. 
As of now, $\mu_{in}$ in sub-10~nm GNRs cannot be extracted from experiments
since line-edge roughness (LER) is presently limiting 
mobility in state-of-the-art GNRs~\cite{Wang,Yang,BettiTED}.   
Theory, on the other hand, allows us to individually evaluate the 
impact of each scattering source on mobility, which is often a prohibitive task in experiments.
Phonon-limited scattering is particularly important, since it represents the
unavoidable scattering mechanism at finite temperature, and provides a good indication of the 
potential for the material in electronics, either for devices or interconnects.

Here we investigate the effect of phonons and surface optical (SO) 
phonons on carrier transport in GNRs by means of a full 
band (FB) approach based on a tight-binding (TB) 
description~\cite{Saito} of the electronic structure and of the 
phonon spectrum. We compute scattering rates with first-order perturbation 
theory using the deformation potential approximation (DPA), and low-field 
mobility using the Kubo-Greenwood formula~\cite{EPAPS}. 
We obtain the one-dimensional (1D) subbands of an armchair GNR from a TB $p_z$-Hamiltonian 
accounting for energy relaxation at the edges~\cite{Son}. 
The electron energy dispersion is quantized in the transverse direction $y$ 
with wavevectors $k_{y\eta}= (2 \pi \eta)/[(l+1)a]$, where $a=$~0.249~nm 
is the graphene lattice constant, 
$l$ is the number of dimer lines and index $\eta$ runs from 1 to $l$. 
Graphene phonon spectrum is obtained with the force constant 
dynamic-matrix approach, including contributions up to the fourth nearest 
neighbors (4NNFC approach)~\cite{Saito} and using force constant parameters 
extracted from first-principles calculations~\cite{Wirtz}. 
Each of the six phonon branches of graphene, 
labelled by the quantum number $j$,  
is splitted into $l$ 1D sub-branches, with transverse 
wave vector $q_{y\beta}$ \{$q_{y\beta}= (2 \pi \beta)/[(l+1)a]$ for 
$\beta=0,...,l/2-1$ and $q_{y\beta}= [2 \pi (\beta+2)]/[(l+1)a]$ for 
$\beta=l/2,...,l-1$ \}. 

The momentum relaxation rate of an electron in the initial state 
${\bf k}=(k_x,k_{y\eta})$ accounting for scattering from GNR 
phonons is obtained from the Fermi Golden Rule, summing over all 
final states ${\bf k'}=(k_x',k_{y\eta'})$, conserving total energy and 
longitudinal momentum~\cite{EPAPS}: 
\begin{eqnarray} \label{eqn:rateTOT}
\!\!\frac{1}{\tau({\bf k})}\!\!&=&\!\!\sum_{\eta'=1}^{l} \sum_{j=1}^6 \!\sum_{\beta}\!  \!\int_{-k_{F}}^{+k_{F}}\!\!\!\!\!\!\!\!\!\!\!dq_x \,\,\frac{n^{\mp}_{\bf q} \, \hbar D_j^2 }{4 \rho W E_{ph}^{j \beta} } \left(1\!+\!\mbox{cos} \,\theta_{\bf{k}\bf{k}'} \right)\left(1\!-\!k_x'/k_x \right) \nonumber \\
&& \!\!\!\!\!\! G_{\eta,\eta',\beta} \, \,\delta \!\left[E\left({\bf k'}\right)\!\!-\!\!E\left({\bf k}\right)\!\mp \!E_{ph}^{j \beta} \left({\bf q}\right)\right] \!\frac{1\!-\!f\left(E_{\bf k'}\right)}{1\!-\!f\left(E_{\bf k} \right)} \, ,
\end{eqnarray} 
where $D_j= q\, D_{AC}$ ($D_j= D_{OP}$), if $j$ is a longitudinal acoustic 
(in-plane optical) mode, $D_{AC}$ and 
$D_{OP}=1.4 \times 10^{11}$~eV/m~\cite{Fang}  are the 
acoustic (AC) and optical (OP) deformation potentials, respectively, 
$k_x'= k_x \pm q_x$, $q=\left| {\bf q} \right|=(q_x^2+q_{y\beta}^2)^{1/2}$, 
$\rho\approx 7.6\times 10^{-8}$~g/cm$^2$ 
is graphene mass per unit area~\cite{Fang}, 
$n_{\bf q}^-$ is the Bose-Einstein occupation 
factor and $n_{\bf q}^+=n_{\bf q}^- +1$. In addition, 
$k_F= \pi/(\sqrt{3}\,a)$, 
$W= a/2\,(l-1)$ is the GNR width, 
$(1+\mbox{cos }\theta_{\bf k k'})$  is the spinor overlap~\cite{Rozhkov}   
and $f(E)$ is the Fermi occupation factor. 
In Eq.~(\ref{eqn:rateTOT}) the upper sign is for 
phonon absorption (ABS) and the lower for phonon emission (EM).
$G_{\eta,\eta',\beta}$ is the form factor due to 
the transverse momentum conservation 
uncertainty~\cite{EPAPS,BettiIEDM10}. 

We have investigated the electrostatic coupling between electrons in the 
GNR channel and remote phonons of the substrate considering 
the GNR 
deposited on an oxide layer of thickness $t_{ox}$, width $W_{ox}$ and placed 
at a distance $d=0.4$~nm~\cite{EPAPS}. 
The oxide is backgated by an ideal metal.
Since phonon modes are almost 
constant as a function of the longitudinal SO phonon 
wavevector~\cite{Konar}, we assume the same energies $E_{SO}^{\beta}$ 
as in Ref.~\cite{Perebeinos} 
for the two considered SO phonon modes. 
In addition, electrons are confined in the plane, 
therefore the electron-SO phonon scattering rate reads~\cite{EPAPS}:
\begin{eqnarray} \label{eqn:rateremote}
\!\!\!\!\!\frac{1}{\tau({\bf k})}\!\!&=&\!\!\!\sum_{\eta'=1}^l \sum_{\beta} \!\! \int_{-k_F}^{+k_F}\!\!\!\!\!\!\!\!\!d Q_x \!\!\int\!\!d Q_y \frac{L W_{ox} e^2 F_{\beta}^2 G \,e^{-2 Q d}}{4 \pi \hbar \,\left[\epsilon_{1D}\left(Q_x\right)\right]^2 Q}\!\left(1\!+\!\mbox{cos} \,\theta_{\bf{k}\bf{k}'} \right) \nonumber \\
&&  n^{\mp}_{\bf Q} \, \delta \left(E_{\bf{k}'} \!-\!E_{\bf{k}} \!\mp \!E_{SO}^{\beta} \right)\!\left(1\!-\!k_x'/k_x \right) \!\frac{1\!\!-\!\!f\!\left(E_{\bf k'}\right)}{1\!\!-\!\!f\!\left(E_{\bf k}\right)} \, ,
\end{eqnarray}
where ${\bf Q} = (Q_x,Q_y)$ represents the 2D SO phonon wavevector, 
$Q=\left|{\bf Q}\right|$, the sum $\sum_{\beta}$ runs over all SO phonon 
modes, $F_{\beta}^2\propto 1/(L W_{ox})$ is the 
electron-phonon coupling parameter~\cite{Konar,EPAPS}, 
$G(\eta,\eta',Q_y)$ is the form factor~\cite{EPAPS} 
which reduces to $G_{\eta,\eta',\beta}$ if $Q_y \!= \!q_{y\beta}$. 

In Eq.~(\ref{eqn:rateremote}), $\epsilon_{1D}\left(Q_x \right)$ 
is the GNR static dielectric function calculated within the Random Phase 
Approximation (RPA) in the size quantum limit~\cite{Fang,EPAPS}. 
For 10~nm-wide GNRs, screening of the electric field due to polar vibrations 
is instead modeled by means of the 2D RPA graphene static dielectric function 
$\epsilon_{2D}\left(Q \right)$~\cite{Hwang}. 
From a numerical point of view, in order to reduce numerical noise, 
we have approximated $\delta$ in Eqs.~(\ref{eqn:rateTOT}) 
and~(\ref{eqn:rateremote}) 
with a Gaussian window of standard deviation $\Delta E$ and, for the lowest AC 
subbranches, a collisional broadening approach 
has been implemented considering 
$\Delta E=\hbar/2 \, \left[1/\tau\left({\bf k},q_{y\beta}\right)\right]$. 
Finally, we have computed the low-field mobility $\mu_{in}$ 
by means of the Kubo-Greenwood formula accounting 
for 1D transport~\cite{EPAPS,BettiIEDM10}. 

DPA formally leads to a zero coupling with the transversal acoustic (TA) and flexural (ZA) phonon modes,
so that only scattering with longitudinal acoustic (LA) modes is typically considered \cite{Fang}.
Theory~\cite{Mariani} and 
Raman spectroscopy~\cite{Ferrari} have shown that ZA modes are negligible down to 130~K. 
However, classical results based on a TB description of 
electron-phonon coupling
\cite{Pietronero} and recent ab-initio calculations~\cite{Borysenko} 
have demonstrated that TA modes play a comparable role as that of LA modes in degrading $\mu_{in}$. 
On the other hand, a physical description of graphene taking into account long-range interaction between carbon atoms
highlights an off-diagonal coupling to the TA modes 
through the modulation of the hopping parameters, which is 
smaller than the on-diagonal deformation potential contribution~\cite{Ando}. 
We choose to adopt a physically consistent approach
and use DPA considering electron coupling only with LA, LO and TO modes, rather than
euristically reintroduce the contribution of TA modes. 
We use $D_{AC} = 10.9$~eV, extracted from DFT calculation for the GNR 
family $3\,l\!+\!1$~\cite{Long}, rather   
than fitting experiments which are affected by uncontrolled mechanisms 
and actually lead to a large spread of the considered values for 
$D_{AC}$~\cite{Fang,Bolotin,Borysenko}.

\begin{figure} [htbp]
\begin{center}
\includegraphics[scale=0.63]{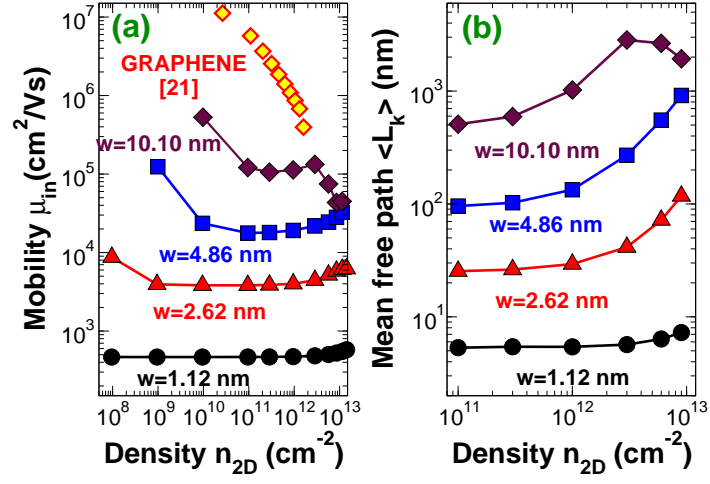}
\end{center}
\caption{(a) $\mu_{in}$ and (b) $\langle L_{\bf k}\rangle$ for 
an electron in the lowest subband as a function of $n_{2D}$ for 
different $W$. DFT calculations~\cite{Borysenko} for 
graphene are also shown in (a).}
\label{fig:totalmu}
\end{figure}

Low-field mobility is shown in Fig.~\ref{fig:totalmu}a as a 
function of the electron density $n_{2D}$ for different widths. 
$\mu_{in}$ close to 500~cm$^2$/Vs is found for 1~nm-wide 
GNR, exceeding by almost one order of 
magnitude the experimental mobility of GNRs~\cite{Wang,Yang} and 
the intrinsic phonon-limited mobility of silicon nanowires~\cite{JinJAP} of comparable size.
We find that $\mu_{in}$ is mainly limited by 
backward scattering involving AC 
phonons, 
due to the large mode-dependent OP energy 
offset~($\approx$~130-160~meV)~\cite{BettiIEDM10}. 
Unlike in 2D graphene, where $\mu_{in}\propto 1/n_{2D}$~\cite{Perebeinos}, 
the lateral confinement in GNRs leads to a non-monotonic $n_{2D}$-dependence 
as also observed in CNTs~\cite{PerebeinosNL2} (Fig.~\ref{fig:totalmu}a). 
For small $W$, $\mu_{in}$ increases with $n_{2D}$, due to the reduction of 
final states available for scattering. For wider GNRs biased in the 
inversion regime, electrons can populate excited subbands opening 
additional channels for scattering, thus reducing $\mu_{in}$. 

\begin{figure} [htbp]
\begin{center}
\includegraphics[scale=0.64]{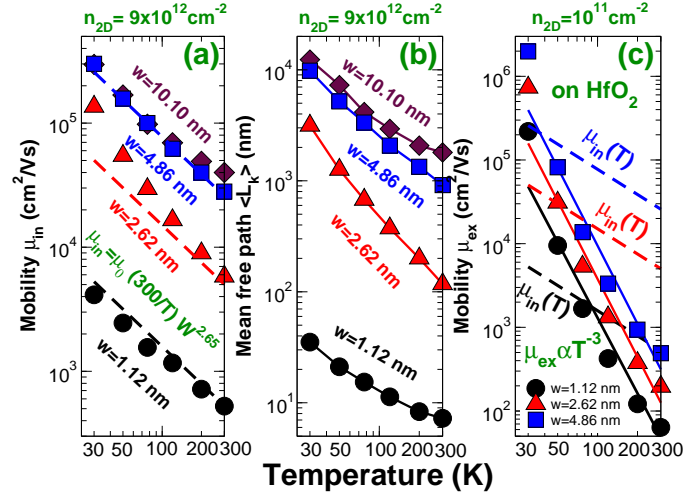}
\end{center}
\caption{(a) mobility $\mu_{in}$, (b) mean free path $\langle L_{\bf k}\rangle$ for the 
lowest subband, and (c) mobility $\mu_{ex}$ 
for GNR on HfO$_2$ as a function of $T$ for different $W$. 
In (a) and (c) dashed lines 
correspond to the empirical formula for $\mu_{in}(W,T)$ and solid lines in (c) 
to the fit for $\mu_{ex}$.} 
\label{fig:mucoherence}
\end{figure}

In Fig.~\ref{fig:mucoherence}a, 
$\mu_{in}$ is plotted as a function of temperature $T$. Similarly to what has been
observed in small-diameter 
CNTs~\cite{Perebeinos2}, in narrow GNRs the dependence on $W$ and $T$ 
can be expressed by means of the empirical relation $\mu_{in}(W,T)= \mu_0 \, (300 \,\mbox{K}/T ) \,(W/1 \, \mbox{nm})^{\alpha_{AC}}$ where 
$\mu_0\approx$~391~cm$^2$/Vs and $\alpha_{AC}=2.65$, 
which is close to $\mu_{in} \propto W^3$ expected for narrow GNRs, since 
$\mu_{in}\propto \tau/(W  \times DOS) \propto G^{-1} DOS^{-2} $ and 
$G \propto 1/W^2$ and the density of states $ DOS \propto 1/\sqrt{W}$. 
Of course, for large $W$ $\mu_{in}$ saturates to that of 2D graphene. 
Since AC in-plane phonons scattering is dominant and 
$n_{\bf q}^{-} \approx kT/\hbar \omega$ for 
$kT \gg \hbar \omega $, $\mu_{in}$ is inversely proportional to $T$ 
(Fig.~\ref{fig:mucoherence}a). 
The mean free path in the first subband $\langle L_{\bf k} \rangle$ 
is shown in 
Fig.~\ref{fig:totalmu}b as a function of $n_{2D}$ and in 
Fig.~\ref{fig:mucoherence}b as a function of $T$, 
where $\langle L_{\bf k} \rangle \equiv
\langle v\left({\bf k}\right) \tau\left({\bf k}\right) \rangle$, 
$v\left({\bf k}\right)$ is the group velocity and the expectation 
value $\langle \cdot \rangle$ has been computed in the Brillouin zone, 
considering $f(1-f)$ as the distribution function~\cite{BettiIEDM10}. 
At $T=300$ K, 
$\langle L_{\bf k} \rangle$ is of the order of few $\mu$m for larger GNRs, 
as expected in graphene flakes, while it is~$\approx$~10~nm 
for narrower GNRs (Figs.~\ref{fig:mucoherence}b). In addition, 
$\langle L_{\bf k} \rangle \propto 1/T$, as $\mu_{in}$ 
(Figs.~\ref{fig:mucoherence}b).

\begin{figure} [htbp]
\begin{center}
\includegraphics[scale=0.75]{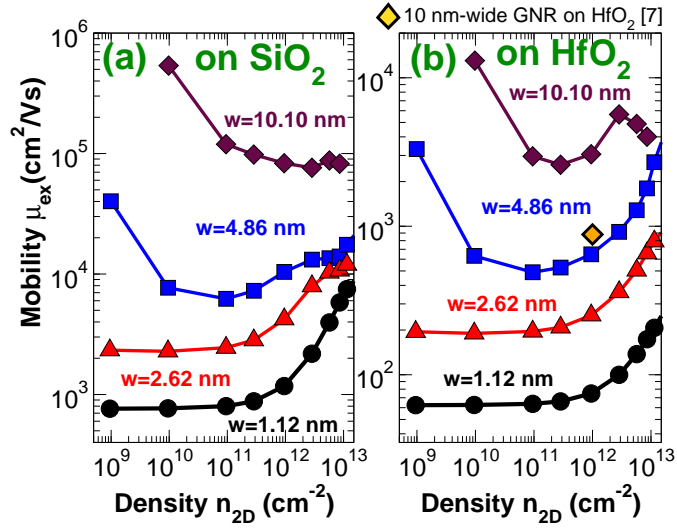}
\end{center}
\caption{$\mu_{ex}$ as a 
function of $n_{2D}$ for GNR deposited (a) on SiO$_2$ and (b) 
on HfO$_2$.}
\label{fig:remote}
\end{figure}

The SO phonon-limited mobility $\mu_{ex}$ as a function of $n_{2D}$ is shown 
in Figs.~\ref{fig:remote}a-b for $W$ smaller than 10~nm, considering GNRs 
deposited both on SiO$_2$ and on HfO$_2$. As in 
graphene~\cite{Konar,Perebeinos}, the higher the dielectric constant, 
the larger the mobility suppression due to SO phonon scattering. 
In particular, we observe $\mu_{ex}$ down to 700 cm$^2$/Vs for SiO$_2$ 
(Fig.~\ref{fig:remote}a) and 60 cm$^2$/Vs for HfO$_2$ 
(Fig.~\ref{fig:remote}b), due to the smaller energy offset of the emission 
processes. 
As in CNTs~\cite{PerebeinosNL}, 
$\mu_{ex}\propto W^{\alpha_{SO}}$ with $\alpha_{SO}$ ($\approx$ 1.4-1.6) 
dependent on $n_{2D}$ and smaller than $\alpha_{AC}$. 
For $W$ smaller than 5~nm 
$\mu_{ex}$ increases with electron concentration due to 
the impact of screening, 
whereas for $W=$~10.10~nm for higher concentrations $n_{2D}$ the increase 
of available modes for scattering reduces mobility. 

Comparing Figs.~\ref{fig:remote}a-b with Fig.~\ref{fig:totalmu}a, it can be 
observed that SO phonons play a secondary role for very narrow GNRs on SiO$_2$ 
but they become predominant with increasing $W\geq$~2.5~nm roughly for 
$n_{2D} \!< \!$~10$^{12}$~cm$^{-2}$, whereas they are predominant 
for all $n_{2D}$ densities in GNRs on HfO$_2$. 
Comparison with experiments shows that
$\mu_{ex}$ is larger by up to one order of magnitude than
mobility measured on GNRs deposited on SiO$_2$~\cite{Wang} 
and by a factor three than mobility measured on 10-nm-wide GNRs integrated with ultrathin 
HfO$_2$ dielectric~\cite{Liao}. This gives a rough estimation
of the increase in mobility that could be achieved through 
fabrication technology improvements capable to suppress
the present dominant scattering mechanisms (e.g. LER). 
As can be noted in Fig.~\ref{fig:mucoherence}c and as it also occurs in CNTs 
deposited on polar dielectrics~\cite{PerebeinosNL}, 
$\mu_{ex} \propto 1/T^{\gamma}$. In particular, 
for HfO$_2$ $\gamma \approx 3$. Since 
$\mu_{in} \propto 1/T$, SO phonon scattering dominates transport 
roughly above 100~K for all $W$ (Fig.~\ref{fig:mucoherence}c), 
as for CNTs on SiO$_2$~\cite{PerebeinosNL}.

\begin{figure} [htbp]
\begin{center}
\includegraphics[scale=0.65]{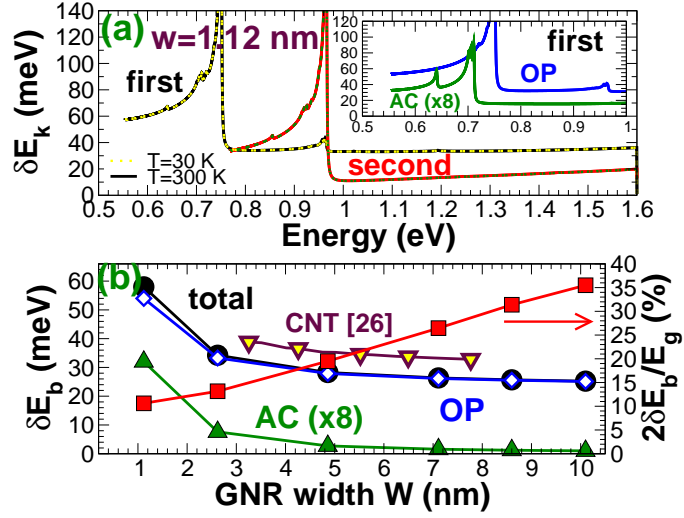}
\end{center}
\caption{(a) Polaronic energy shift $\delta E_{\bf k}$ as 
a function of the energy $E_{\bf k}$ for the first two subbands 
($W=$~1.12~nm). Solid curves correspond to $T\!=$~300~K, dotted curves to 
$T\!=$~30~K. Inset: OP and AC ($\times$~8) contributions to 
$\delta E_{\bf k}$ for the first subband. (b): Polaronic binding energy 
$\delta E_b$ (left) and polaronic correction to the band gap 
$2 \delta E_b/ E_g$ (right) as a function of $W$. Results for 
CNTs~\cite{Perebeinos2} are also reported.} 
\label{fig:binding}
\end{figure}

Finally, we focus on the polaronic energy shift $\delta E_{\bf k}$ 
due to the electron-phonon coupling, computed exploiting the 
second-order perturbation theory~\cite{EPAPS}. 
Fig.~\ref{fig:binding}a shows $\delta E_{\bf k}$ 
as a function of $E_{\bf k}$ for the lowest two subbands for the 
$W=1.12$~nm case. $\delta E_{\bf k}$ 
is weakly energy dependent near the cutoff subband, is independent of $T$ 
and increases sharply in correspondence of intersubband transitions. 
As in CNTs~\cite{Perebeinos2}, mostly OP phonons contribute to 
$\delta E_{\bf k}$ (inset of Fig.~\ref{fig:binding}a). 
Instead, unlike in CNTs~\cite{Perebeinos2}, the contribution to 
$\delta E_{\bf k}$ from AC phonons exhibits few peaks due to the transverse 
momentum conservation uncertainty (inset of Fig.~\ref{fig:binding}a). 
The polaronic binding energy, i.e. the polaronic energy shift referred to the 
first conduction subband edge $E_{C1}$, is 
$\delta E_b\!\!= \!\!\delta E_{\bf k} (E_{\bf k}\!\!=
\!\!E_{C1})$ and is almost~59~meV for 1~nm-wide GNRs, 
close to that obtained for semiconducting CNTs with the same number 
$l$~\cite{Perebeinos2} (Fig.~\ref{fig:binding}b) 
and corresponds to a band gap renormalization 
$2 \, \delta E_b \approx$~118~meV for 1-nm nanoribbons,
and to a relative correction of -35\% of the energy gap $E_g$ 
of 10-nm nanoribbons (Fig.~\ref{fig:binding}b). 

In conclusion, we have proposed a very accurate full-band approach to evaluate 
low-field phonon-limited mobility $\mu_{in}$ 
in GNRs. We find that 
$\mu_{in}$ is close to 500~cm$^2$/Vs 
in suspended 1~nm-wide GNRs at room temperature, and is suppressed
down to 60~cm$^2$/Vs in 1-nm wide GNR deposited on HfO$_2$, due to coupling
with SO phonons. The result is important from the point of view of methodology and of 
fundamental physics, since the corresponding mean free paths range from 1 to 
10~nm, undermining the possibility of performing ballistic or coherent transport experiments at non-cryogenic temperatures. It is also important from the application point of view:
whereas suspended 2D graphene has an intrinsic mobility at room temperature several orders of magnitude larger than that of bulk semiconductors, narrow GNRs with reasonable semiconducting gap have only slightly larger mobility than comparable silicon nanowires. 
Finally, we also find polaron formation in armchair GNRs, with a 
remarkable band gap renormalization of up to 35\% in the case of 10~nm-wide ribbons. 

This work was supported in part by the EC 7FP 
through NANOSIL (n.~216171), GRAND (n.~215752) grants, 
and by the MIUR-PRIN project GRANFET (Prot. 2008S2CLJ9).
Authors thank www.nanohub.org for the provided computational resources.

\newpage

\newpage

\newpage

\newpage

\newpage

\newpage


\end{document}